\documentstyle[aps,tighten,graphicx,prl,twocolumn]{revtex}

\begin{document}
\title{The Einstein-Podolsky-Rosen 
Paradox and Entanglement 2: Application to Proof of Security for 
Continuous Variable Quantum 
Cryptography\\}
 \vskip 1 truecm
\author{M. D. Reid\\}
\address{Physics Department, 
The University of Queensland,Brisbane,
Australia}
\date{\today}
\maketitle
\vskip 1 truecm
\begin{abstract}
In a previous paper  certain measurable criteria have been derived, that are 
sufficient to demonstrate the existence of Einstein-Podolsky-Rosen (EPR) 
correlations for measurements with continuous variable outcomes. Here it 
is shown how such EPR criteria, which do not demand perfect EPR correlations,  
 can be used to prove the extent of security for continuous variable quantum 
 cryptographic schemes (in analogy to that proposed by Ekert) where Alice and 
Bob hope to construct a secure sequence of values from measurements performed 
on continuous-variable EPR-correlated fields sent from a distant source. It 
is proven that the demonstration of the EPR criterion on  Alice's and Bob's 
joint statistics compels a necessary loss in the ability to infer the results 
shared by Alice and Bob, by measurements performed on any third channel 
potentially representing an eavesdropper (Eve). This result makes no 
assumption about the nature of the quantum source of the fields transmitted 
to Alice and Bob, except that the EPR correlations are observed at the 
final detector locations. In this way  a means is provided to establish security
 in the presence of some loss and less than optimal correlation, and against 
any  eavesdropping strategy employed by Eve prior to detection of the fields 
by Alice and Bob.
 
\end{abstract}
\narrowtext
\vskip 0.5 truecm
\section{Introduction}

Einstein, Podolsky and 
Rosen~\cite{epr} (EPR) presented a now famous argument in 1935 in an attempt to 
show that quantum mechanics is an ``incomplete'' theory. Their 
argument was based on the premise of ``no action-at-a-distance'' and 
made 
assumptions about the nature of ``reality''. In 1966 Bell~\cite{Bell} showed 
 that 
the predictions of all theories (called local hidden variable 
theories) consistent with these EPR premises would
 satisfy certain constraints called Bell inequalities. He also showed that 
for some situations the predictions of quantum mechanics will violate 
these Bell inequalities, meaning an incompatibility of quantum 
mechanics with local hidden variables. 

While Bell's original work, and subsequent experiments relating to it, 
applied to situation of discrete spin   
measurements, the original EPR argument was presented for ``position'' 
and ``momentum'' measurements 
with  continuous 
variable  outcomes. The experimental observation of such continuous variable 
EPR correlations have been 
achieved~\cite{ou,zhang,silber} 
  using fields, where the 
 conjugate ``position'' and 
 ``momentum'' observables are replaced by   
 the two orthogonal noncommuting 
 quadrature phase amplitudes of the field~\cite{eprquad}. 
  The theoretical proposal~\cite{eprr,rd} relating to these experiments employed a 
 two-mode squeezed state~\cite{cavessch} as the source of EPR fields. For such 
 experiments it is not possible to demonstrate the perfect 
 correlation as discussed originally by EPR. The experimental 
 signatures are based on a criterion first presented~\cite{eprr} in 1989, and 
 expanded on in a recent paper~\cite{mdrentcrit}. 
     The EPR fields generated through the two-mode squeezing 
     interaction have enabled the experimental 
  realization of a
   continuous variable quantum teleportation~\cite{tele}.

Quantum cryptography using squeezed or two-mode squeezed states 
predicting EPR correlations for  
quadrature phase amplitude measurements with continuous variable 
outcomes have been recently 
investigated~\cite{cryralph,cryhill,cryreid,crycerf,cryhjk,crynavz,crysil,cryduan}.
 Of particular interest here 
 is the continuous variable quantum cryptographic 
scheme~\cite{cryreid,cryhjk,crynavz} analogous to that 
discussed by Ekert~\cite{numbercry} 
for spin-$1/2$ systems where Alice and Bob wish to construct a secure 
key from correlated  data sent to each of their locations from an 
 entangled continuous variable EPR source. 
 The original proposal of Ekert 
proposed to use the correlated spin state shown by Bohm~\cite{Bohm} to 
demonstrate a 
 version of the EPR paradox relating to measurements with discrete 
 outcomes. 
Bell showed in 1966 that this state (the Bell-state) 
violates a Bell inequality, and in 
Ekert's proposal the violation of the Bell 
inequality is used to demonstrate security.

The direct continuous variable 
``position /momentum'' measurements 
that demonstrate the EPR paradox for the two-mode squeezed state 
cannot by any simple 
rotation of measurement angle demonstrate 
a violation of a Bell inequality. The point of this paper is to 
emphasize that this does not however diminish the  
usefulness of such a state in for example providing secure mechanisms for quantum 
cryptography protocols, since one can replace the Bell-inequality used 
in Ekert's protocol by 
an EPR-criterion to test for security.  

In this paper we prove how the {\it demonstration of EPR correlations, 
using 
 the 1989 EPR criterion, by Alice and Bob on their two detected 
 channels puts a limit 
on the accuracy of any inference made by Eve, about the results of the 
measurements performed by Alice and Bob.}  
Importantly this is proved {\it for 
any quantum source}, meaning security against any strategies Eve is 
able to  
employ prior to Alice and Bob detecting the fields.

	To summarize the conclusions of this paper, it is shown that the 
	determination by Alice and Bob of a perfect, maximum EPR 
	correlation  in their 
	detected fields implies security against  any hypothetical
	 Eve  obtaining the key 
	sequence. If Eve has 
	intercepted {\it one} or {\it both} of the EPR 
	channels (from the source to Bob 
	or from the source to Alice) in any manner, to obtain the key sequence with 
	any degree of accuracy (to give a noninfinite variance in her estimate of the 
	values), then it is proved that the 
	EPR correlation detected by Alice and Bob could no longer be maximum.  
	It has also been proved that there is no alternative 
	set of quantum fields (source) available to Eve, that would enable 
	her  to obtain the key sequence with any degree 
	of accuracy, and still retain the optimal EPR correlation measurable between Bob and 
	Alice.
	
	The situation where a reduced EPR correlation is observed between 
	Alice and Bob is more subtle. Alice and Bob would expect a 
	certain degree of EPR correlation based on measurements 
	performed on their EPR 
	source (and perhaps an 
	expected degree of loss on transmission). If their measured EPR 
	correlation is noticeably reduced on one  particular 
	transmission then this could 
	indicate Eve's presence (through extra loss) on Bob's channel,
	 and certainly it would be wise to 
	retransmit. However {\it even in such a situation where eavesdropping 
	might 
	have occurred}, (or in the situation 
	where the EPR correlation is apparently as expected but where it 
	cannot be excluded  
	that Eve has 
	substituted a different quantum source), our procedure 
	{\it allows a deduction of the minimum degree of fuzziness 
	of Eve's estimate of the key values shared by Alice and Bob}. 
	
	The 
	strategy here is to  derive limits on Eve's knowledge of the key sequence, 
	based only on the reliable measurement of a certain amount of EPR 
	correlation between Alice and Bob. This result is of current relevance in that 
	fixed amounts of EPR correlation for continuous variable outcomes 
	have been (irrefutably) confirmed experimentally 
	(whereas Bell inequality violations have not).
	 The proposal is to then use these limits 
	to encode the  message in such a way as to elude 
 Eve.

 \section{EPR criteria based on conditional measurements}
 
 We first need to define the EPR criteria, and here the 
 results of a previous paper~\cite{eprr,mdrentcrit} are summarized. 
 Consider two quantum fields 
   propagating towards two spatially separated location 
   at $A$ and $B$ respectively. The fields will be generated by a 
   appropriate quantum source so that the results of 
   certain measurements are correlated. 
   Two 
 observables $\hat{x}$ (the ``position'') 
 and $\hat{p}$ (``momentum'') are defined for the subsystem at location 
 $A$. These observables satisfy an 
 uncertainty relation 
 \begin{equation}
 \Delta \hat{x} \Delta \hat{p}\geq C \label{eqn:hur}
 \end{equation}
but where we will consider from this point on that with 
  appropriate scaling the 
$\hat{x}$ and $\hat{p}$ are now dimensionless and $C=1$. 
A measurement  $\hat{x}^{B}$ made at $B$ gives a result $x_{i}^{B}$.  
In this paper, $i$ is used to label the possible 
   results, discrete or otherwise, of the measurement $\hat{x}^{B}$. 
   The results of measurements $\hat{x}$ and $\hat{x}^{B}$
    at $A$ and $B$ are correlated, so 
   that the measurement at $B$ enables a prediction to be made 
   about the result of a measurement $\hat{x}$ at $A$. 
  We define a set 
  of distributions $P(x|x_{i}^{B}¥)$ giving the probability of a result 
  $x$ for the
    measurement at $A$, conditional on a result $x_{i}^{B}¥$ for measurement 
    at $B$. The variance and mean of the 
    conditional distribution $P(x|x_{i}^{B}¥)$ are denoted by  
    $\Delta_{i}^{2}x$ and $\mu_{i}$ respectively.

  Also, for certain correlated fields, we can infer the result of  
 measurement $\hat{p}$ at $A$, based on a measurement,
  $\hat{p}^{B}¥$ say, at $B$. We denote the results of the measurement 
  $\hat{p}^{B}¥$ 
  at $B$ by $p_{j}^{B}¥$. We also define the 
 probability distribution,  $P(p|p_{j}^{B}¥)$ for obtaining the 
 result $p$ upon 
  measurement of $\hat{p}$ at $A$, 
conditional on the result $p_{j}^{B}¥$ 
 for the measurement $\hat{p}^{B}¥$ at $B$. 
   The variance of the conditional  distribution $P(p|p_{j}^{B})$ 
  is denoted by $\Delta_{j}^{2}p$. 

 The situation discussed by Einstein-Podolsky-Rosen demands a 
 perfect correlation between the result of measurements $\hat{x}$ at 
 $A$ and $\hat{x}^{B}$, and also between $\hat{p}$ at $A$ and 
 $\hat{p}^{B}$ at $B$. For this case, the variances of the 
 conditional distributions  must satisfy 
 \begin{equation}
  \Delta_{i}x= \Delta_{j}p=0
\end{equation}
 for all $i,j$. 
  This situation however is not achievable for 
continuous variable measurements. 

It has been discussed in the previous paper~\cite{mdrentcrit} how 
EPR correlations would be demonstrated 
where one can establish that each of the conditional 
distributions $P(x|x_{i}^{B}¥)$
  is very narrow, 
so that 
\begin{equation}
P(x|x_{i}^{B}¥)=0\quad,where\quad  |x-\mu_{i}|>\delta \label{eqn:narrow}
\end{equation}
 and $\delta<1$. A similar result must 
be proved for each 
$P(p|p_{j}^{B}¥)$. As discussed previously~\cite{mdrentcrit}
this situation represents the spirit of the original EPR 
gedanken experiment in its truest form, but is difficult to achieve 
experimentally.

     A more readily achieved criterion still sufficient to demonstrate EPR 
    correlations was proposed in 
 1989~\cite{eprr} and has been further explained in the previous 
 paper~\cite{mdrentcrit}. 
 We first define the   
 weighted variance
 \begin{eqnarray} 
 \Delta_{inf}^{2}\hat{x}=\sum_{i}P(x_i^{B}¥)\Delta_{i}¥^{2}x
 \end{eqnarray} 
 and similarly
 \begin{equation}
  \Delta_{inf}^{2}\hat{p}=\sum_{j}P(p_j^{B}¥)\Delta_{j}¥^{2}p
  \end{equation}
 Here $P(x_{i}^{B})$ is the probability for a 
 result $x_{i}^{B}$ upon measurement of $\hat{x}^{B}$, and $P(p_{j})$ is 
 defined similarly. It has been shown that the observation of 
 \begin{equation}
\Delta_{inf}\hat{x}\Delta_{inf}\hat{p}< 1 \label{eqn:eprcrit}
\end{equation}
 implies a demonstration of EPR correlations (the EPR 
paradox).

It is mentioned that the evaluation of the conditional distributions 
 for each outcome of the continuous variable $x_{i}^{B}¥$ at $B$ 
 is not always be practical. 
 It has been discussed~\cite{eprr,mdrentcrit} previously how it 
 is possible to perform other measurements, closely 
 related to squeezing measurements, that are also  sufficient to 
 indicate the EPR/entangled nature of the system. This is the approach 
 used experimentally to date to demonstrate EPR correlations.
  We propose 
 upon a result $x_{i}^{B}¥$ for the measurement at $B$ that the predicted 
 or estimated value for the result $x$ at $A$ is given  by the linear estimate  
 $x_{est}=gx_{i}^{B}¥+d$. 
 The  
deviation $\delta_{i}¥=x-(gx_{i}^{B}¥+d)$ in this linear estimate can 
in principle then
 be measured, and an average over the different values of $x_{i}^{B}¥$ 
 evaluated. We obtain as a measure 
of average error in our inference of the result at $A$, based on the 
result at $B$, and using this linear approach:
\begin{eqnarray}
\Delta_{inf,L}^{2}\hat{x}&=&\sum_{x_{i}^{B}¥}P(x_{i}^{B}¥)
\langle\delta^{2}\rangle_{i}\nonumber\\
&=&\langle\{\hat{x}-(g\hat{x}^{B}¥+d)\}^{2}\rangle.
\end{eqnarray}
 The best linear estimate $x_{est}$ is the one that will minimize  
$\Delta^{2}_{inf,L}\hat{x}$.  
The best choice for $g$ is discussed in~\cite{eprr}. 
Where $x_{est}=\mu_{i}¥$ it follows that  
the variance $\Delta^{2}_{inf,L}\hat{x}=\Delta_{inf}^{2}x$. 
Generally however $\Delta^{2}_{inf,L}\hat{x}\geq\Delta_{inf}$. 
 The observation of
   \begin{equation}  
    \Delta_{inf,L}\hat{x}\Delta_{inf,L}\hat{p}< 
    1\label{eqn:eprcritlin}
    \end{equation}
      implies
    quantum  inseparability, 
        for any $g$ and $d$, and also 
        the situation of the EPR paradox.

\section{The two-mode squeezed state as the quantum EPR source}

  Suppose the two quantum fields are generated via the 
 interaction Hamiltonian  
 $H_I = i\hbar \kappa(\hat{a}^{\dagger} \hat{b}^{\dagger} 
 -\hat{a}\hat{b})$, where $\hat{a}$ and $\hat{b}$ symbolize the boson 
 operators 
 for the fields at $A$ and $B$ respectively. 
  For vacuum initial states this 
 interaction generates, after a finite time $t$,
  two-mode squeezed light~\cite{cavessch} 
\begin{eqnarray}
	|\psi> = \sum_{n=0}^{\infty}
	c_{n}¥|{n}>_{a} |{n}>_{b} \label{eqn:twomode}
	\end{eqnarray}
 where $c_{n}= tanh^{n}r/cosh r$ and $r=\kappa t$. This interaction 
  provides a quantum model for the parametric amplifier. 
  
  This simple quantum 
 state was shown to be EPR-correlated in reference~\cite{eprr}, 
and EPR correlations using parametric interactions and the 
criteria (\ref{eqn:eprcritlin}) 
have been achieved experimentally. We define the quadrature phase amplitudes 
 \begin{eqnarray}
		\hat{x}&=&\hat{X}_a =(\hat{a}+\hat{a}^\dagger)\nonumber\\
		\hat{p}&=&\hat{P}_a= (\hat{a}-\hat{a}^\dagger)/i\nonumber\\ 
		\hat{x}^{B}&=&\hat{X}_b =(\hat{b}+\hat{b}^\dagger)\nonumber\\
		\hat{p}^{B}&=&\hat{P}_b= 
		(\hat{b}-\hat{b}^\dagger)/i\label{eqn:quad}
		\end{eqnarray}
that are measurable using local oscillator and homodyne detection 
techniques that were developed originally in efforts to generate and detect 
squeezed light. The Heisenberg 
uncertainty relation is 
 $\Delta^{2}X_{a}¥\Delta^{2}P_{a}¥\geq  1$. 
 
 It is seen from the linear EPR criterion (\ref{eqn:eprcritlin}) that the following is also 
 a criterion sufficent to demonstrate EPR correlations in the spirit 
 of the original EPR paradox:
 \begin{equation}
      \Delta(X_{a}-gX_{b})\Delta(P_{a}+hP_{b})<1 \label{eqn:strong}
\end{equation}
where we have used the linear form 
\begin{equation}
\Delta_{inf}x=\Delta_{inf,L}^{2}¥\hat{x}=
\langle(X_{a}-gX_{b})^{2}¥\rangle\label{eqn:defineinf}
\end{equation}  
and $\Delta_{inf}^{2}¥p=\Delta_{inf,L}^{2}¥\hat{p}=\langle(P_{a}+hP_{b})^{2}¥\rangle$, 
and $g$ and $h$ are parameters chosen to minimise the variances (the 
choice of $d$ is best at $0$ since the quadrature amplitude means for the vacuum squeezed 
state are zero).
  
 The two-mode squeezed state  
 predicts 
~\cite{eprr} ($g=h=\tanh 2r$) the correlations 
$X_{a}=X_{b}$, and $P_{a}=-P_{b}$ to give 
\begin{eqnarray}
	\Delta_{inf,L}^{2}\hat{x}=\Delta_{inf,L}^{2}\hat{p}=1/\cosh 
	2r,
\end{eqnarray}
a clear demonstration of EPR correlations for all $r$.

 \section{The Crypotographic Scheme}
 
We now consider the application of the EPR state to quantum 
cryptography with continuous variable outcomes. 
 To summarize, an EPR source 
emits two fields, one which propagates towards 
Alice at location $A$, and the other to Bob at location $B$. 
As one possible strategy, 
Alice selects to measure randomly either quadrature phase amplitude 
$X_{a}=\hat{x}$, 
corresponding to angle $\theta=0$, or quadrature phase amplitude 
$P_{a}=\hat{p}$, corresponding to angle 
choice $\theta=\pi/2$, say. Similarly Bob will measure randomly either 
quadrature phase amplitude $X_{b}=\hat{x}^{B}$, 
corresponding to angle $\phi=0$, or quadrature phase amplitude 
$P_{b}=\hat{p}^{B}$, corresponding to angle choice $\phi=\pi/2$. As 
discussed in~\cite{mdrentcrit}, and shown previously in~\cite{eprr}, 
for the choices 
$\theta=\phi=0$, as the two-mode squeeze parameter $r$ becomes 
large, the results $x$ of Alice's measurement   
and $x_{i}^{B}$ for Bob's measurement, will become 
identical. For the choices $\theta=\phi=\pi/2$ 
the results $p$ for Alice and $p_{j}^{B}$ are also correlated 
(anticorrelated in fact): 
for large $r$ we have $p=-p_{j}^{B}¥$.

In the style of the original quantum cryptographic proposals~\cite{numbercry},
we consider here the proposal where Alice communicates to Bob after the 
measurements (through a public channel) her choice of measurement angle 
$\theta$ and the result for the measurement, for a 
subensemble, randomly selected after the detections. 
 Bob is able to check  his measurements and compare his value 
 for the result of measurements that should be correlated with Alice's.

Alice and Bob can then use their shared subensemble to 
calculate the conditional probabilities $P(x|x_{i}^{B}¥)$ and 
$P(p|p_{j}^{B}¥)$ 
and the associated respective variances $\Delta_{i}^{2}¥x$, $\Delta_{j}^{2}¥p$ of these 
distributions defined in Section 2.  Here $x$ is Alice's 
result for $\hat{x}$, and $x_{i}^{B}¥$ is the result for Bob's measurement 
(which we have symbolized by $\hat{x}^{B}$) 
correlated with $\hat{x}$. Similarly $p$ is Alice's result for 
$\hat{p}$, and $p_{j}^{B}¥$ is the result for Bob's measurement (symbolized 
by $\hat{p}^{B}$) anticorrelated 
with $\hat{p}$. For our particular two-mode squeezed state 
(\ref{eqn:twomode}) the prediction is
\begin{eqnarray}
\Delta_{i}x=\Delta_{j}p=1/\surd\cosh 
	2r
 \end{eqnarray}
for all $i,j$. 

The   
maintenance of the EPR correlation 
(between Alice and Bob's fields) is  determined through 
examination of the individual $\Delta_{i}x$ or $\Delta_{j}p$, and  
through a measured degree of  
violation of the 1989 EPR criteria 
(\ref{eqn:eprcrit}). The proposal is that this is used to check, 
or establish a degree of, security.

 For example if perfect EPR 
correlations are established between Bob and Alice, it will be proved 
that there could have been no intervention on 
the channels from the source to Alice and Bob, or 
reconstruction of an alternative source, by an eavesdropper Eve.
Where Alice and Bob are able to confirm  narrow conditional distributions 
satisfying (\ref{eqn:narrow}), it is possible to establish the 
necessity of a certain 
degree of fuzziness in Eve's data.
 Once security is established, the 
measurement angle for the remaining subset is shared, and where the 
choice of angle is to predict correlation between results, the 
sequence of common shared 
values can be used in some manner to form a key.

\section{Proof of Security}

We now need to give the proof of security for such an EPR scheme. 
It is assumed 
as usual that the choice of angles $\theta$ and $\phi$ 
(whether to measure $\hat{x}$ or 
$\hat{p}$) for Bob's and Alice's measurements are 
randomly and independently chosen after the transmission of the 
fields to Alice and Bob at secure spatially separated locations. Alice 
and Bob make  
delayed-choice measurements. 
Therefore we assume that an 
eavesdropper (Eve) cannot anticipate the angle choice prior to Alice 
and Bob receiving the fields. This must also be true of the selection of 
the subensemble used by Alice and Bob to evaluate the statistics to 
test security. In 
this way it is assumed that the statistics evaluated by Bob and Alice 
on the
subensemble accurately reflect the 
statistics of the entire fields received by Alice and Bob. 

It is the objective here to prove security against {\it any 
strategy Eve could adopt prior to the detection of the transmitted fields by 
Alice and Bob}. 
For example Eve might 
 interfere with  and retransmit {\it one or both} of the 
 fields in some manner to forward to Alice and/ or  Bob. Alternatively  
she might {\it sabotage the EPR source} to substitute 
an alternative three-channeled correlated source, where the 
three transmitted beams propagate to Alice, Bob and also Eve at a third 
location.  Eve could then potentially perform a final measurement after 
public communication of Alice's angle choice. Alternatively, where 
Alice and Bob use an EPR source with less than optimal correlations, 
as would be the realistic situation, Eve might 
replace the EPR source with one showing  
improved EPR correlations. This might enable her to tap some of the signal 
for the purpose of eavesdropping, while the decrease in EPR 
correlations that could be a consequence of her tapping would go 
unnoticed by Alice and Bob who expect a more weakly correlated signal 
anyway.  

It is possible to consider cryptographic schemes where security is 
established on the basis of 
the assumption of a secure source, and also a 
secure channel from the source to 
Alice (see for example~\cite{crynavz}). In such schemes, the 
tapping of the channel from the source to Bob shows as a loss of EPR 
correlations which can be detected by Alice and Bob, alerting them to 
Eve's interference.

 However for systems of  
perfect EPR correlation (such as generated from the Bohm-Bell state 
used in the original scheme of Ekert) 
a stronger proof of security is possible without these assumptions of secure 
source and second EPR channel. It then becomes relevant 
 to determine the extent of such security possible for the continuous 
 variable two-mode squeezed EPR state (\ref{eqn:twomode}), whose EPR 
 correlations have 
 been experimentally confirmed, but 
 which for finite squeeze parameter $r$ is always less 
 than optimally correlated.

  \begin{figure}
   \includegraphics[scale=.55]{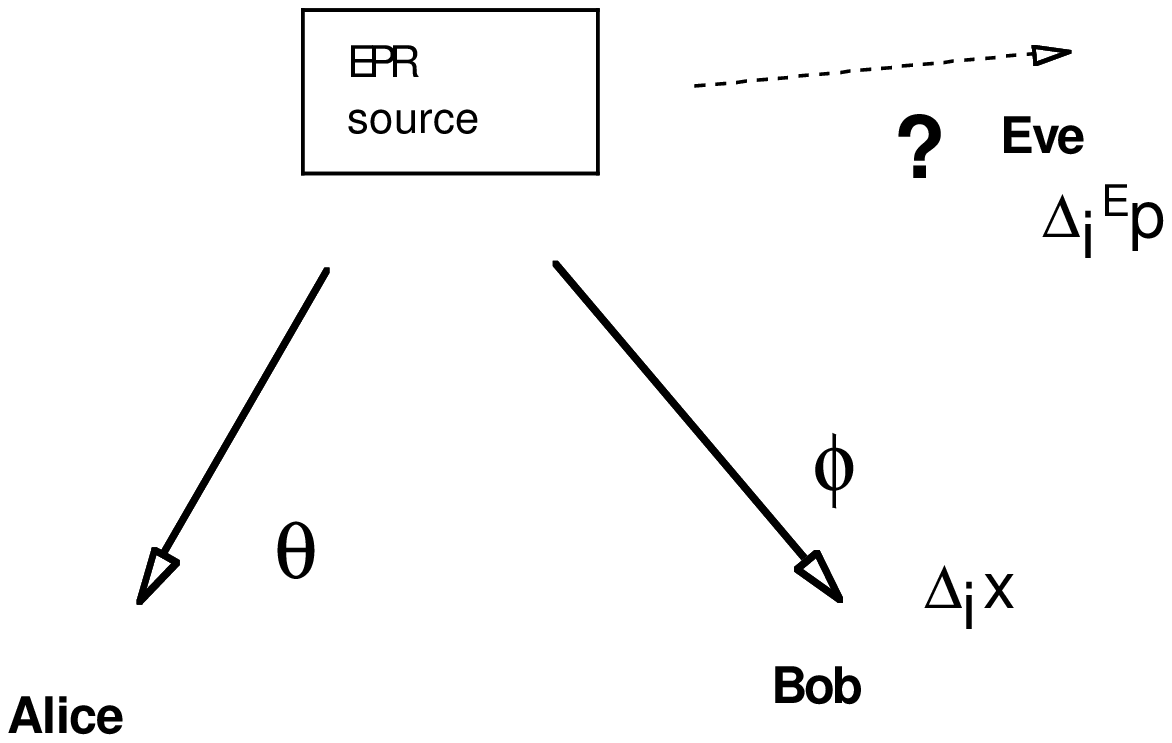}
\caption[fig2]
{EPR-correlated fields reach Alice and Bob who 
perform delayed-choice measurements (``position'' or ``momentum'') 
as determined by 
the choice of $\theta$ or $\phi$) at spatially separated locations.
 The correlated  results 
  can then  form 
the shared values for a key. If Alice and Bob can prove that 
their fields are EPR-correlated according to the 1989 criterion 
(\ref{eqn:eprcrit}), then it is impossible for any  
eavesdropper Eve at a third location to have an accurate 
 replica of the correlated results shared by Alice and Bob. Eve's 
 estimate of Alice's values are necessarily less accurate than 
 Bob's, and the degree of inaccuracy can be evaluated from the 
 details of the EPR-correlation. This is true 
regardless of the nature of the original quantum source.}%
\end{figure}

The proof of security presented here then 
gives a method {\it to determine the level of 
security based only 
on the nature  of the measured statistics evaluated between Alice and Bob after 
the detection of their fields}, and therefore involves {\it 
no assumptions regarding 
the amount of loss} occurring during propagation or the 
{\it nature of the original quantum state}.  
 We present a proof of security by  demonstrating 
 the impossibility of Eve, an 
eavesdropper at a third spatial location, 
having (or being able to obtain) a perfect copy of the  
 results $x,p$ shared by Alice and Bob, if Alice and Bob 
  measure through 
their subensemble an  EPR 
 correlation based on their measurements $\Delta_{i}x$ and $\Delta_{j}p$ 
 (see Figure 1). 

An eavesdropping process by Eve results in the generation of a final 
quantum state describable by a density operator symbolized by 
$\rho$. For example if Eve attempts to extract information by 
intercepting Bob's channel, quantum mechanically Eve's measurement 
process is 
represented by a Hamiltonian that acts for some duration, there being 
an initial quantum state describing   
 Bob's and Eve's (and Alice's) fields. 
The final state $\rho$
that is produced after the interaction (we may also  
consider a series of interactions that may involve destructive 
measurements and 
state generation) describes Alice, Bob and Eve's final fields that are 
 eventually detected and undergo 
measurements by 
Alice, Bob and Eve at their different final spatial locations. 
 
Eve attempts to gain the results of Alice's $\hat{x}$ (or 
$\hat{p}$) 
through some measurement on her field  
symbolized by $\hat{x}^{E}$ (and $\hat{p}^{E}$).
The  quantum state $\rho$ 
predicts probability distributions for the outcomes of all possible 
measurements performed by Alice, Bob and Eve: for example a 
probability distribution  
$P_{x,x,p}¥(x,x_{i}^{B},p_{q}^{E}¥)$ for the  outcomes $x,x_{i}^{B},p_{q}^{E}$ 
of Alice's, Bob's and Eve's results of measurement 
$\hat{x},\hat{x}^{B},\hat{p}^{E}$; 
and also a $P_{p,x,p}(p,x_{i}^{B}¥,p_{q}^{E})$ for the  outcomes 
$p,x_{i}^{B}¥,p_{q}^{E}$ 
of Alice's, Bob's and Eve's results of measurement 
$\hat{p},\hat{x}^{B},\hat{p}^{E}$.

We define 
the 
probability $P(x|\{x_{i}^{B}¥,p_{q}^{E}¥\})$ of a result $x$ for Alice's measurement of 
 $\hat{x}$, conditional on the results $x_{i}^{B}¥$ and 
 $p_{q}^{E}¥$  for Bob's $\hat{x}^{B}¥$ 
and Eve's  $\hat{p}^{E}¥$ respectively. We also define the 
probability $P(p|\{x_{i}^{B}¥,p_{q}^{E}\})$ of a 
result $p$ for Alice's measurement of 
 $\hat{p}$, given  the results $x_{i}^{B}¥$ and 
 $p_{q}^{E}¥$ for Bob's $\hat{x}^{B}¥$ 
and Eve's  $\hat{p}^{E}¥$ respectively. 

A constraint is placed on the variances 
$\Delta_{i,q}^{2}x$ and $\Delta_{i,q}^{2}p$ of the conditional distribution 
$P(x|\{x_{i}^{B}¥,p_{q}^{E}¥\})$ and 
$P(p|\{x_{i}^{B}¥,p_{q}^{E}\})$ respectively, 
 for any possible quantum state 
$\rho$.
The predicted statistics of 
Alice's measurements conditional on measurements performed by Bob and 
Eve are described by the reduced density operator $\rho_{A}=\langle 
x_{i}^{B}¥|\langle 
p_{q}^{E}¥|\rho|p_{q}^{E}¥\rangle|x_{i}^{B}¥\rangle/N$ (where $N$ is a 
normalization factor).
 The variance $\Delta_{i,q}^{2}x$ of the conditional distribution 
$P(x|\{x_{i}^{B}¥,p_{q}^{E}¥\})$ gives the uncertainty in the 
estimate of Alice's $\hat{x}$ conditional on 
the  results $x_{i}^{B}$ and  $p_{q}^{E}¥$ for Bob's $\hat{x}^{B}¥$ 
and Eve's  $\hat{p}^{E}¥$  measurements. 
 Bob's (and Eve's) measurement constitutes a measurement 
of Alice's $\hat{x}$, with precision $\Delta_{i,q}x$.
 The uncertainty relation will imply 
the constraint (for the quadratures as defined by (\ref{eqn:quad}), $C=1$)
\begin{equation}
\Delta_{i,q}p \geq 1 /\Delta_{i,q}x \label{eqn:uncertaintycrp}
\end{equation}

The marginal distribution $P(x|x_{i}^{B})$, the probability of 
 Alice's result $x$ for measurement $\hat{x}$ conditional on the 
 result $x_{i}^{B}$ for Bob's measurement $\hat{x}^{B}$, is given by 
 \begin{eqnarray}
P(x|x_{i}^{B})&=&\sum_{p_{q}^{E}}P_{x,x,p}(x,x_{i}^{B},p_{q}^{E})/P(x_{i}^{B}) 
\nonumber\\
&=&\sum_{p_{q}^{E}} 
P(x|\{x_{i}^{B},p_{q}^{E}\})¥f_{p_{q}^{E}} \label{eqn:margprobcrp}
\end{eqnarray} 
where $f_{p_{q}^{E}}$ is the fraction
 $f_{p_{q}^{E}}=P(x_{i}^{B},p_{q}^{E})/P(x_{i}^{B})$; 
$P(x_{i}^{B},p_{q}^{E})$ is the probability for result $x_{i}^{B}$ 
and $p_{q}^{E}$, respectively,
 upon joint measurement $\hat{x}^{B}$ and $\hat{p}^{E}$; and 
 $P(x_{i}^{B})$ is the probability of $x_{i}^{B}$ for Bob's 
 measurement of $\hat{x}^{B}$. 
This implies the following relationship for the variance 
$\Delta_{i}^{2}¥x$ of the 
conditional distribution $P(x|x_{i}^{B})$.
\begin{equation}
\Delta_{i}^{2}¥x \geq \sum_{p_{q}^{E}}f_{p_{q}^{E}}\Delta_{i,q}^{2}¥x
\label{eqn:averagecrp}
\end{equation}

The accuracy of the information obtainable by Eve is determined by the 
standard deviation $\Delta^{E}_{q}p$ of $P^{E}¥(p|p_{q}^{E})$, the 
conditional 
distribution for result $p$ for Alice's measurement of $\hat{p}$ given a 
result $p_{q}^{E}¥$ for Eve's $\hat{p}^{E}$. This marginal distribution is given by
 \begin{equation}
P^{E}¥(p|p_{q}^{E})=\sum_{x_{i}^{B}} 
P(p|\{x_{i}^{B},p_{q}^{E}\})¥f_{x_{i}^{B}} \label{eqn:probpmargcrp}
\end{equation} 
where the fraction $f_{x_{i}^{B}}$ is defined as
$f_{x_{i}^{B}}=P(x_{i}^{B},p_{q}^{E})/P(p_{q}^{E})$; $P(p_{q}^{E})$ is 
the probability for Eve's result $p_{q}^{E}$ upon measurement of 
$\hat{p}^{E}$. 
The  $\Delta^{E}_{q}p$ is related to an average of these variances as 
given by 
\begin{equation}
(\Delta_{q}^{E})^{2}\geq \sum_{x_{i}^{B}}f_{x_{i}^{B}}\Delta_{i,q}^{2}p
\label{eqn:averagecrpsec}
\end{equation}

\subsection{The case of perfect correlation}

First we present a proof of security for the case where Alice and Bob 
measure perfect EPR 
correlations, meaning that all
$\Delta_{i}x=\Delta_{j}p=0$.
 This is the case if the quantum source is a simultaneous 
eigenstate 
of $X_{a}¥-X_{b}¥$ and 
 $P_{a}¥+P_{b}¥$, which is closest to the situation originally defined by EPR.  
 The  
variance $\Delta_{i}^{2}¥x$ is zero in this case we consider 
initially 
of perfect EPR correlations. It 
must follow therefore from (\ref{eqn:averagecrp}) 
that each $\Delta_{i,q}x$ is also zero. Using the constraint 
(\ref{eqn:uncertaintycrp}) each 
$\Delta_{i,q}¥p$ must then be infinite.
Therefore each $\Delta^{E}_{q}p$ 
must also be infinite.
 Each conditional variance, for all 
possible outcomes $p_{q}^{E}$, is infinite, meaning that any   
measurement performed by Eve will give an 
infinite uncertainty in the prediction of Alice's $p$. 

The same logic applied to joint measurements of Bob's $\hat{p}^{B}$ and 
Eve's $\hat{x}^{E}$ implies an infinite 
variance for Eve's estimate of Alice's $x$. In this way it is deduced 
that Eve's estimates of each of Alice and Bob's sequential $x,p$ 
values (these constitute the final key) will have an 
associated infinite uncertainty. 

	To summarize, the determination by Alice and Bob of an optimal EPR 
	correlation  
	\begin{equation}
	\Delta_{i}x=0,\Delta_{j}p=0
	\end{equation}
	 for all $i,j$, in their 
	detected fields implies security that there can be no hypothetical
	 Eve, at a third location as indicated in Figure 1, able to obtain the key 
	sequence. This is proof that if Eve has 
	intercepted {\it one or both the EPR channels} 
 to obtain the key sequence with 
	any degree of accuracy (so that there is a 
	noninfinite variance in her estimate of the Alice and Bob's key 
	values), the 
	EPR correlation would necessarily have been reduced to give a nonzero 
	result for at least one of the $\Delta_{i}x$, $\Delta_{j}p$. 
	It has also been proved that there is no alternative 
	set of quantum fields (source) available to Eve, that would  
	allow her to obtain the key sequence with any degree 
	of accuracy and still retain the optimal EPR correlation between Bob and 
	Alice.

 \subsection{The case of reduced correlation but where all conditional 
  distributions are narrow}
 
Considering that a practical experiment will not 
have perfect correlation (values of $\Delta_{i}x$ might typically be 
$0.7$ for current situations~\cite{silber}), we need to argue more generally. 
	It is still possible to derive   
	limits on Eve's knowledge of the key sequence, 
	based {\it only on the reliable measurement of a certain amount of EPR 
	correlation between Alice and Bob} (and the assumption that Eve does not 
	have access to the random choice of 
	Alice and Bob's measurement angles and subensemble 
	selection, both of which are selected after transmission and 
	detection of the fields by Alice and Bob). 
	 The proposal, such as that discussed in Section 6, is to use these limits 
	to encode the  message in such a way as to defeat Eve.
	
Based on Alice and Bob's measurements over the subensemble, the 
conditional probability distribution $P(x|x_{i}^{B})$ (and 
 $P(p|p_{j}^{B})$) can be measured by Alice 
and Bob, and their associated 
variances $\Delta_{i}^{2}¥x$ (and $\Delta_{j}^{2}¥p$) can be evaluated. 
We examine in this subsection the case where all conditional distributions 
measured by Alice and Bob are shown to have a nonzero but small standard 
deviation, so that for example where these distributions are Gaussian  
Alice and Bob demonstrate $\Delta_{i}x<1/3$ and $\Delta_{j}p<1/3$.

Most 
generally the variances such as $\Delta_{i}^{2}¥x$ are 
related to the individual variances $\Delta_{i,q}^{2}¥x$ 
  by the relation (\ref{eqn:averagecrp}). 
 Although the $\Delta_{i}x$ might be 
small, an individual $\Delta_{i,q}$ might  not be. The 
possibility cannot be ruled out that
 Eve is able to obtain upon some (small) fraction of her 
 measurements the result of 
Alice's $p$ to good accuracy. 
  The relationship given by (\ref{eqn:margprobcrp}) is
 certainly true however. 
 Suppose all  (that is for all $i$) 
 the conditional distributions $P(x|x_{i}^{B})$ measured by Alice and Bob are
 sufficiently narrow so that 
 the probability of obtaining a result $x$ outside a range 
 $\mu_{i}-\delta\leq x 
 \leq \mu_{i}+\delta$ is zero ie
\begin{equation}
P(x|x_{i}^{B}¥)=0\quad,\quad for\quad|x-\mu_{i}|>\delta
\end{equation}
 and we assume $\delta<1$ ($C=1$). A similar result must 
be proved for each 
$P(p|p_{j}^{B}¥)$.
  Recall here $\mu_{i}$ is the mean of the particular 
 distribution $P(x|x_{i}^{B})$.
 In this case based on (\ref{eqn:margprobcrp}),
  we can say for sure that, for a given fixed $x_{i}^{B}$, 
 each of the $P(x|\{x_{i}^{B},p_{q}^{E}\})$ must 
 also satisfy $P(x|\{x_{i}^{B},p_{q}^{E}\})=0$ outside the range 
 $\mu_{i}¥\pm\delta$. 
 This implies 
 that each variance $\Delta_{i,q}^{2}¥x$ (of 
 $P(x|\{x_{i}^{B},p_{q}^{E}\})$) could not exceed the value of $\delta^{2}¥$, 
 implying in turn by (\ref{eqn:uncertaintycrp}) 
 that each $\Delta_{i,q}p$ must satisfy 
 \begin{equation}
 \Delta_{i,q}p\geq 1/\delta \label{eqn:limit}.
 \end{equation}
 In this way, since this is true for all $i$, and 
 using (\ref{eqn:averagecrpsec}), it 
 is proved that the uncertainty (standard deviation) in {\it each} of Eve's 
 estimates of Alice's $p$ (this uncertainty is the standard deviation 
 of the conditional 
 distribution $P(p|p_{q}^{E})$ as defined above) will exceed $1/\delta$.
  \begin{equation}
 \Delta_{q}^{E}¥p\geq 1/\delta \label{eqn:limitfour}.
 \end{equation}

 For 
 the two-mode EPR state (\ref{eqn:twomode}), the conditional 
 distributions are predicted to be Gaussian with variance given as 
 $\Delta_{i}x=\Delta_{j}p=1/\surd\cosh 
2r$. Of course the actual distributions must be measured by Alice and 
Bob as part of the security procedure.  
For $r$ sufficiently large  so that (recall $\delta\leq 1$)
\begin{equation}
 \Delta_{i}x < \delta/3 \leq 1/3 \label{eqn:limittwo}
\end{equation}
 the Gaussian distribution is predicted to be 
negligible at $x>\delta$, and 
there is then proof that Eve's best possible estimates satisfy 
\begin{equation} 
\Delta^{E}_{q}p \geq 1/\delta. \label{eqn:limitthree}
\end{equation}
This means that as Eve performs the measurement $\hat{p}^{E}$ to obtain a result 
$p_{q}^{E}$, the standard deviation of the conditional distribution 
$P^{E}(p|p_{q}^{E})$ (for Alice's result $p$ conditional on Eve's outcome)
 exceeds $1/\delta$ {\it for 
every possible outcome} $p_{q}^{E}$, and for all possible 
measurements $\hat{p}^{E}$.  Similar logic applied to Alice and Bob's 
conditional distributions $P(p|p_{j}^{B})$ gives a corresponding limit on the 
 error in each of Eve's estimates of Alice's result for $\hat{x}$. 
 
 In 
 this way it is derived that a minimum degree of uncertainty or 
 fuzziness (as given by $\Delta^{E}_{q}p \geq 1/\delta $, $\Delta^{E}_{q}p \geq 
 1/\delta$) exists for Eve's estimate of every piece of the key 
 sequence shared by Alice and Bob. If $\eta_{Eve}$ is Eve's estimate 
 of the particular key value, and $\eta$ is Alice's actual key 
 value, then we have $\langle(\eta_{Eve}-\eta)^{2}¥\rangle^{1/2}¥ \geq 1/\delta$.
   
\section{A possible encryption protocol}

	Using 
the prediction for the two-mode squeezed state 
  $\Delta_{inf}\hat{x}=\Delta_{inf}¥\hat{p}
  =\Delta_{j}\hat{p}=\Delta_{i}\hat{x}=1/\surd\cosh 
	2r \rightarrow 0$ as $r\rightarrow \infty$, we see 
	that as the squeeze parameter $r=\kappa t$ is increased it 
	becomes possible for Bob to resolve Alice's $x$ or $p$ value while 
	Eve can only resolve with error $\rightarrow 
	\infty$. The 
	measured quadrature phase amplitude values shared between Alice and 
	Bob form a  secure key sequence denoted by the sequence
	$\eta_{m}¥$, $m=1,2,\ldots$. We will define the key $\eta_{m}$ to 
	consist of Alice's relevant measured values $x$ (or 
	$p$), though in the 
	limit of $r\rightarrow\infty$   
 there is no deviation of Bob's measured values $x_{i}^{B}¥$ (or 
	$p_{j}^{B}$) from Alice's.
	
	The data given by 
	the variable $z_{m}¥$ is encoded, suppose simplistically to give a transmitted
	classical amplitude or number 
	$y_{m,sent}¥=z_{m}¥+A\eta_{m}¥$ (where $A$ is a relative 
	amplification factor). The key $\eta_{m}$ 
	known to both Alice and Bob enables Bob to decode 
	the signal $z_{m}$, whereas 
	Eve will have an  
	infinite uncertainty in her measurement of $\eta_{m}¥$, and 
	therefore $z_{m}$.
	
	 The chief 
	difficulty for Alice and Bob comes for finite $r$ where the EPR 
	correlation is reduced.  Suppose initially that 
	Alice and Bob's measurements of the EPR 
	correlation and associated conditional distributions, 
	enable them  to establish that the probability distribution 
	$P(x|x_{i}^{B})$ of Alice's result $x$ conditional 
	on Bob's result $x_{i}^{B}$  is
	a distribution  
	with mean $\mu_{i}¥$ and 
	standard deviation $\sigma$. The two-mode squeezed state 
	(\ref{eqn:twomode}) predicts a Gaussian distribution
	$\mu_{i}¥=\tanh rx_{i}^{B}$ for $\theta=0$ 	
	and $\sigma=1/\surd\cosh 
	2r$. We suppose a similar result is achieved 
	for  $P(p|p_{j}^{B})$: (\ref{eqn:twomode}) predicts the Gaussian with
	$\mu_{j}¥=-\tanh rp_{j}^{B}¥$  	
	and $\sigma=1/\surd\cosh 
	2\kappa t$. 
	
	Bob's key sequence is the sequence 
	$\eta_{m,Bob}$ that he builds up by selecting, for each of his 
	relevant measurements, 
	 $\eta_{m,Bob}=\mu_{i}$ where he obtained an outcome $x_{i}^{B}$ upon 
	 measurement of $\hat{x}^{B}$, or $\eta_{m,Bob}=\mu_{j}$ where he 
	obtained $p_{j}^{B}$ upon measurement of $\hat{p}^{B}$. 
	The deviation of Bob's key value from Alice's key value is then
	\begin{equation}
	\langle(\eta_{m}¥-\eta_{m,Bob})^{2}\rangle=\sigma
	\end{equation}
	(The choice $\eta_{m,Bob}=\mu_{i}, \mu_{j}¥$ made by Bob minimizes this 
	rms error.)
	Bob's estimate of the decoded data $z_{m}¥$ is 
	$z_{m,Bob}=y_{m,sent}-A\eta_{m,Bob}$. His rms error is
	\begin{eqnarray} 
	\langle 
	(z_{m}¥-z_{m,Bob})^{2}\rangle
	&=& A^{2}\langle(\eta_{m}¥-\eta_{m,Bob})^{2}\rangle \nonumber\\
	&=& A^{2}\sigma^{2}.
	\end{eqnarray}
 A satisfactory binning by Alice 
	of her  data $z_{m}¥$ enable Alice and Bob to share precisely such a signal 
	sent by Alice. This is determined by Alice and Bob, based on their 
	knowledge of the conditional statistics measured over the subensemble. 
	For example let us assume that Alice and Bob's distributions are 
	Gaussian. In this case there is a negligible chance (.0027) 
	of $|\eta_{m}¥-\eta_{m,Bob}|$
	being greater than $3\sigma$, so that if Alice restricts the
	 $z_{m}¥$ to be one of a series of numbers separated by $6A\sigma$, then Bob 
	will round off correctly to reconstruct the correct signal. 
	
	  However Eve's decoded data is  $z_{m,E}=y_{m,sent}-A\eta_{m,Eve}$ 
	where $\eta_{m,Eve}¥$ is Eve's key. Consider the situation discussed above 
	in equation (\ref{eqn:limittwo}) where every one of Bob's 
	measured conditionals are Gaussian and satisfy 
	$\Delta_{i}x=\Delta_{j}p=\sigma=\delta/3$ ($\delta<1$). 
	With (\ref{eqn:limitthree}) we conclude 
	that each standard deviation  
	of Eve's conditional distributions $P^{E}¥(x|x_{r}^{E})$ and 
	$P^{E}¥(p|p_{q}^{E})$ 
	satisfies 
	$\Delta_{q}^{E}p>1/\delta $, $\Delta_{r}^{E}x>1/ \delta$. The 
 rms error of Eve's signal must satisfy ($\sigma=\delta/3$)
		\begin{eqnarray} 
	\langle 
	(z_{m}¥-z_{m,Eve})^{2}\rangle&=&A^{2}\langle(\eta_{m}¥-\eta_{m,Eve})^{2}\rangle
	\nonumber\\
	&\geq&A^{2}/\delta^{2}=A^{2}/9\sigma^{2}¥
	\end{eqnarray}
 Suppose Bob and Alice's correlation reveals 
$\sigma=1/3$ (which is the largest value that is sensible to this 
particular approach). Then Eve's 
best could not do better than $\langle(z_{m}¥-z_{m,Eve})^{2}\rangle^{1/2}¥= A$.

	On the basis of the assumption of a particular form for 
	 Eve's conditional distributions $P(x|x_{q}^{E})$ and $P(p|p_{r}^{E})$ 
	 (eg Gaussian), 
	a minimum error rate for Eve's information could be calculated. The 
	probability of Eve evaluating Alice's $z_{m}$ outside of the 
	range $z_{m}\pm 3A\sigma$, and to therefore establish the incorrect 
	value for $z$, is significant if Eve's conditional distributions are 
	Gaussian (the probability of an incorrect $z_{m,Eve}$ being $0.3173$). 
	The Gaussian calculation for Eve's error rate is relevant, in that 
	the  EPR channels generated from the source (\ref{eqn:twomode}), 
	and subsequently interfered with by Eve through mechanisms 
	able to be modeled by linear 
	interaction Hamiltonians/couplings 
	such as 
	($\hat{a_{Eve}¥}$ and $\hat{b}$ symbolize the boson 
 operators 
 for Eve's and Bob's fields respectively)
	\begin{equation}
	H_I = \kappa(\hat{a}_{Eve}^{\dagger} \hat{b}^{\dagger} 
 +\hat{a}_{Eve}\hat{b})\label{eqn:lossyint}
 \end{equation}
  with vacuum or squeezed state inputs, 
	would predict such Gaussian conditional distributions. Examples of 
	such linear eavesdropping strategies, have been discussed 
	previously~\cite{cryralph,cryhill,cryreid,crycerf}.
		
It is noted however that while a limit is placed on the variances of 
 Eve's conditionals, the form of these distributions has not been 
 constrained by the very general approach presented here. For example 
 Eve's eavesdropping might produce an 
 $\eta_{m,Eve}¥$ that has a
	significant probability peak located at or near zero, 
	meaning that a significant 
	number of the $z_{m}¥$ will be  read with no error. 
	Alice and Bob would need to use 
	the fact that then necessarily other measurements performed by Eve 
	must have very significant deviation from the true $z_{m}¥$ value
	 as part of the encoding scheme. A more sophisticated scheme is 
	 required, as for 
	 example one based on a
	 scrambling of data where every value of the $z_{m}¥$ is needed 
	 to descramble completely, and where significant single errors in a 
	 $z_{m}¥$ or $\eta_{m}$ value compound Eve's reduced  
	 ability to decode.

\section{Proof of security for weaker correlation}

  The above protocol requires narrow conditional distributions, 
 $\Delta_{i}x,\Delta_{j}p<1/3$ (ie $\sigma<1/3$ for Gaussian 
 distributions).  With reported measured values of
 $\Delta_{inf}^{2}¥x=\sum_{i}P(x_{i})\Delta_{i}^{2}¥x\approx 
 0.7$ such a value is probably not currently 
 achievable. Here we present a more general strategy which can apply 
 where variances satisfy $\Delta_{i}x<.57$, $\Delta_{j}p<.57$.

 We define the set of probabilities 
 $\{P(x_{r}^{E}¥), P(p_{q}^{E}¥)\}$ and uncertainties 
 $\{\Delta^{E}_{r}x,\Delta^{E}_{q}p\}$ which determine 
 the 
  accuracy of Eve's information on the key sequence (this being 
 Alice's sequence of relevant $\hat{x}$,$\hat{p}$ results). 
 Here $P(x_{r}^{E}¥)$ is the 
 probability of Eve obtaining a result $x_{r}^{E}$ upon a 
 measurement $\hat{x}^{E}$, and $\Delta^{E}_{r}x$ is the standard deviation 
 of the probability distribution $P^{E}(x|x_{r}^{E})$ for Alice's 
 result $x$ for $\hat{x}$, conditional on Eve's result $x_{r}^{E}$ 
 for $\hat{x}^{E}$. This set must be compared 
 with the set of probabilities $\{P(x_{i}^{B}¥),P(p_{j}^{B}¥)\}$ and 
 uncertainties 
 $\{\Delta_{i}x,\Delta_{j}p\}$ that determine Bob's 
 accuracy of information of Alice's $x,p$ data.
 
  First, where the correlation 
 between Alice's and Bob's data is sufficient to satisfy the 1989 
 criteria 
 (\ref{eqn:eprcrit}) for EPR, it can be shown that the sets of statistics are 
 necessarily different: that determining Eve's information involving 
 greater uncertainties than that determining Bob's information.
 We show this as follows.

 If we assume that Eve's set is identical to Bob's, we obtain a 
 contradiction. 
 We could define the joint probability $P_{i,q}=P(x_{i}^{B},p_{q}^{E})$ 
 of the  
 result $x_{i}^{B}¥$ for Bob's measurement $\hat{x}^{B}$ and of the result 
 $p^{E}_{q}$ for Eve's measurement  $\hat{p}^{E}¥$
As before we define the variances $\Delta_{i,q}^{2}x$ and 
$\Delta_{i,q}^{2}p$ of the probability distributions for Alice's 
result of measurement $\hat{x}$ and $\hat{p}$ respectively, 
conditional on Bob's and Eve's results 
$x_{i}^{B}¥$ 
 $p^{E}_{q}$. 
 The
  prediction for the average 
  conditional variance (as measured by Alice and Bob) is given by  
  $\Delta_{inf}^{2}\hat{x}=
  \sum_{i}P(x_{i}^{B}¥)\Delta_{i}¥^{2}x$.
 Also if Eve's inferred statistics are to be the same as Bob's, 
 the quantity  $\Delta^{E}_{inf}¥^{2}¥\hat{p}
=  \sum_q P(p_{q}^{E}¥)(\Delta_{q}^{E}p)^{2}¥$
 measured on Eve's statistics 
must equal Alice and Bob's measure of the average $\Delta_{inf}^{2}p$. 
 Applying (\ref{eqn:averagecrp}) and (\ref{eqn:averagecrpsec}) 
 and the Cauchy-Schwarz 
  inequality we would always predict
  \begin{eqnarray}
     \Delta_{inf}^{2}\hat{x}\Delta_{inf}^{2}\hat{p}&=& 
     \Delta_{inf}^{2}\hat{x}\Delta^{E}_{inf}¥^{2}\hat{p} \nonumber\\
     &=&(\sum_{i}P(x_{i}^{B}¥)\Delta_{i}¥^{2}x)
     (\sum_q P(p_{q}^{E}¥)(\Delta_{q}^{E}p)^{2}¥) \nonumber\\
   &\geq & (\sum_{i,q}P(x_{i}^{B}¥)f_{p_{q}^{E}¥}\Delta_{i,q}^{2}x)
     (\sum_{i,q}¥P(p_{q}^{E}¥)f_{x_{i}^{B}¥}\Delta_{i,q}^{2}p)\nonumber\\
     &=&(\sum_{i,q}¥P_{i,q}¥\Delta_{i,q}^{2}x)
     (\sum_{i,q}¥P_{i,q}¥\Delta_{i,q}^{2}p)\nonumber\\
     &=&\langle\Delta_{i,q}¥^{2}x\rangle\langle\Delta_{i,q}¥^{2}p\rangle \nonumber\\ 
    &\geq& |\langle \Delta_{i,q}x \Delta_{i,q}^{E}p \rangle |^{2}\geq 
    1\label{eqn:eprproof}
    \end{eqnarray}
 This is not the case given that  
   Bob's and Alice's statistics show the 1989 EPR criterion.
  In other words, the demonstration of the 
 EPR criteria for Bob's and Alice's statistics ensures 
  that there is a loss of information, as compared to Bob, on Eve's channel.

An increase in Eve's error of inference on the data shared by 
Alice and Bob follows necessarily from Alice and Bob's measurements 
of the general EPR correlations using the 1989 criterion. It is  
required however to 
 employ this fact in a satisfactory way to enable Bob full information 
on a signal transmitted by Alice, while leaving Eve unable to decode.  
Above we have considered strategies where all conditionals have narrow 
variances ($\Delta_{i}x<.3$, $\Delta_{j}p<.3$ for all $i,j$) in 
the fashion of a strong EPR paradox.
	
	Now we consider particular strategies for the situation of inference 
	variances  $\Delta_{inf}x> .3$. First it is 
	possible to prove that 
	\begin{equation}
	\Delta_{inf}^{E}p\geq 1/\Delta_{inf}x\label{eqn:infeqn}
	\end{equation}
	From result (\ref{eqn:averagecrpsec}) we have
	\begin{eqnarray}
	(\Delta_{inf}^{E})^{2}p¥&=&\sum_{q}P(p_{q}^{E})(\Delta_{q}^{E})^{2}p\nonumber\\
	&\geq&\sum_{i,q}P(x_{i}^{B},p_{q}^{E})\Delta_{i,q}^{2}p
	\end{eqnarray}
		\begin{eqnarray}
	\Delta_{inf}^{2}x¥&=&\sum_{i}P(x_{i}^{B})\Delta_{i}^{2}x\nonumber\\
	&\geq&\sum_{i,q}P(x_{i}^{B},p_{q}^{E})\Delta_{i,q}^{2}x
	\end{eqnarray}
Using the Cauchy-Schwarz inequality and (\ref{eqn:uncertaintycrp}) 
we obtain (\ref{eqn:infeqn}).

	Let us suppose then that Alice and Bob establish a uniform set of 
	Gaussian conditional distributions with variances
	\begin{equation}
	\Delta_{i}x=\Delta_{j}¥p=\Delta_{inf}x=\sigma<1.
	\end{equation}
	Alice can choose to bin her signal values to the nearest number of a 
	sequence separated by $6A\sigma$, as described in Section 6.
	 For the case where the two-mode 
	squeezed state is used as a source the conditional distributions are 
	Gaussian (this must be verified by Alice and Bob upon measurements). 
	The probability of Bob decoding the wrong signal value (this is the 
	probability that his value for the key deviates from Alice's by more 
	than three standard deviations) is 
	therefore negligible (.0027), meaning that 
	Bob can use his slightly fuzzy key to decode correct signal values. 
	
	 The 
	 average deviation of Eve's estimate (the mean of her 
	conditional distribution) from Alice's measured key value $x$ (or $p$) 
	is 
	given by the average variances of her conditional distributions. 
	These must satisfy
	\begin{eqnarray}
	\Delta_{inf}^{E}p&\geq& 1/\sigma¥\nonumber\\
	\Delta_{inf}^{E}x&\geq& 1/\sigma\label{eqn:restrictave}.
	\end{eqnarray}
The probability of an error (that Eve will decode Alice's signal $z_{m}$ 
incorrectly) is the probability that Eve's conditional distributions 
($P^{E}(x/x_{r}^{E})$, $P^{E}(p/p_{q}^{E})$)  
deviate from the mean by an amount greater than $3\sigma$. 

	First, provided $1/\sigma>3\sigma$ ($\sigma<.57$),
	 it is necessary that Eve, 
in order to achieve an average variance 
	$\Delta_{inf}^{E}p$ or $\Delta_{inf}^{E}x$ satisfying 
	(\ref{eqn:restrictave}),
will have key values deviating from Alice's key value by an amount 
greater than $3\sigma$: $|\eta_{m}-\eta_{m,Eve}|>3\sigma$ for some 
value of $\eta_{m,Eve}$. This is true {\it for any hypothetical 
  eavesdropping 
scheme Eve might have employed}. In other words it is proven that Eve will 
decode at some point  to obtain wrong signal value 
$z_{m}$ sent by Alice. 
Of course 
	the signal values $z_{m}$ that are now shared accurately by Alice and 
	Bob, but not by Eve, 
	need not form the final message, but can be used as a discrete key to 
	encode a further signal.  
	
	 A calculation of the 
	Eve's error rate based on the assumption that her 
	conditionals are Gaussian distributions with equal 
	$\Delta_{q}^{E}$ is however immediately possible, for any $\delta<1$. For 
	example, where  
	$\sigma=\Delta_{i}x=\Delta_{j}¥p=\Delta_{inf}x=\Delta_{inf}p=.57$, 
	the probability of Eve's error is  $.32$.
	
	Of course as discussed in Section 6, for absolutely secure 
	cryptography, the exact nature of Eve's 
	conditional distributions cannot be measured by Alice and Bob and 
	therefore cannot be assumed. Since in 
	(\ref{eqn:restrictave}) we only 
	restrict the average inference  error, we have not ruled 
	out that Eve is able to  
	 achieve very narrow conditional distributions for most $q$, to obtain the 
	correct result for Alice's signal for most of the signal sequence. 
	This situation however could only be achieved 
	 if Eve has a very significant 
	$\Delta_{q}x$,  
	for some $q$, and therefore a high deviation 
	between Alice's and Eve's measured key values for some 
	of the key sequence, which would cause a large deviation 	$z_{m,Eve}-z_{m}¥$ 
	of Eve's decoded signal from 
  Alice's, for some $m$.  As discussed in Section 6, 
  the encoding protocol would then need to make use of, not only
  an Eve's 
  error rate, but of possibly large individual errors, to reduce her 
  ability to decipher any final message.
	
	\section{Conclusion}

It is proved in this paper that fields demonstrated by Alice and 
Bob (at 
two spatially separated locations) to have certain EPR correlations, 
enable Alice and Bob to share the results of measurements to a great
accuracy. This accurate knowledge of a sequence of results of certain 
measurements cannot be shared by  a 
 third experimenter or eavesdropper  Eve at a different location. 
 In the case of EPR correlations that are less than ideal, certain
 limits on Eve's accuracy of inference of the sequence of values 
 that  form the key have been derived. This conclusion  makes 
 no assumptions about loss or the nature of the quantum source, 
 except that EPR correlations are measured by Alice and Bob, and 
 therefore provides a security against all strategies Eve may take 
 to eavesdrop prior to Alice and Bob receiving the fields.
 
For situations where EPR correlations between Bob and Alice are 
not perfect, it is still possible for Alice and Bob to reconstruct a shared 
key or signal sequence where the values are shared with perfect accuracy. 
The fuzziness placed on Eve's key values means 
that Eve will necessarily at some point 
decipher incorrectly, and some specific strategies are presented for 
the case where 
the averages of the variances of 
Bob's conditional distributions (determining Alice's result based on 
his measurements) satisfy $\Delta_{inf}x<.57$ and $\Delta_{inf}p<.57$. 
Specific 
error rates for Eve's key or decoded signal can be established where a 
particular form (for example a Gaussian as would be 
the case for various linear eavesdropping strategies) for her conditional distributions 
are assumed. Generally, the given encoding scheme must use 
Eve's proven nonzero error rate for a key sequence to  
 establish that it is not  practical for Eve to  
decipher a final message.  

Lastly a comment must be made on what could be concluded on the basis 
of measurements that would appear to be currently achievable 
($\Delta_{inf}x<.57$ probably is not). Reported measurements are 
close to $\Delta_{inf}x=\Delta_{inf}p=.7$. The inference variances are 
measured experimentally in this case through the linear estimate described 
by (\ref{eqn:defineinf}). Our results 
(\ref{eqn:infeqn}) 
 then prove that Eve's conditionals 
satisfy $\Delta_{inf}^{E}x>1.4$, $\Delta_{inf}^{E}p>1.4$.
Assuming that the conditionals 
could be measured to be Gaussian and uniform with a standard 
deviation  
$\sigma=\Delta_{i}x=\Delta_{j}p=.7$, we could apply the strategy discussed 
in Section 6 to allow Bob to share Alice's discretized data without significant 
error. The probability 
of Bob making an error is the probability that the conditional 
distribution gives 
greater than $3\sigma=2.1$ (a negligible error rate of 
.0027).

 If we assume Eve's conditional distributions are also 
Gaussian, then her standard deviation is at least $1.4$ and 
 her error rate is at least $0.136$, fifty times greater than Bob's. 
This gives a proven measure of a level of security against all 
eavesdropping strategies employed by Eve that would result in her 
conditionals being Gaussian. Such strategies include the use of any 
two-mode squeezed state-EPR source of the type (\ref{eqn:twomode}) generated by 
the parametric 
amplification discussed in Section 3, in conjunction with any lossy 
mechanisms or 
eavesdropping strategies involving linear beamsplitters such as given 
by the coupling (\ref{eqn:lossyint}). We cannot prove this generally however 
for any strategy taken by Eve, since she may have a conditional 
distribution with standard deviation $1.4$ but where none of her results 
deviate from the mean by more than $2.1$
 (to give the same error rate on Alice's 
discretized data as Bob's). It is noted in conclusion however that it 
has been proved generally (see equation (\ref{eqn:eprproof})) 
that Eve's estimates of the  
continuous values that form Alice's original key sequence 
 are more fuzzy than Bob's.


\begin{references}
\small
%
 \bibitem{epr} A. Einstein, B. Podolsky and N. Rosen, Phys. Rev. {\bf 47}, 777, (1935).
%
\bibitem{Bell}	J. S. Bell, Physics, {\bf 1}, 195, (1965). 
%
\bibitem{ou}
Z. Y. Ou, S. F. Pereira, H. J. Kimble and K. C. Peng, 
Phys. Rev. Lett. {\bf 68}, 3663 (1992).
%
\bibitem{zhang} Yun Zhang, Hai Wang, 
Xiaoying Li,Jietai Jing, Changde Xie and Kunchi 
Peng, Phys. Rev. A {\bf 62}, 023813 (2000).
%
\bibitem{silber} Ch. Silberhorn, P. K. Lam, O. Weiss, F. Koenig,
  N. Korolkova and G. Leuchs, 
Phys. Rev. Lett. {\bf 86}, 4267 (20001).
%
 \bibitem{eprquad} M. D. Reid and P. D. Drummond, Phys. Rev. Lett. 60, 2731, 
 (1988). P. Grangier, M. J. Potasek and B. Yurke, Phys. Rev. A {\bf 
38}, 3132, (1988). B. J. Oliver and C. R. Stroud, Phys. Lett. A {\bf 
135}, 407, (1989).
%
 \bibitem{eprr} M. D. Reid, Phys. Rev. A {\bf 40}, 913 (1989).
 %
 \bibitem{rd}   M. D. Reid and P.D. Drummond, Phys. Rev. A40, 4493 (1989).
 P. D. Drummond and M. D. Reid,  Phys. Rev. A41, 3930 (1990).
%
\bibitem{cavessch} C. M. Caves and B. L. Schumaker, Phys. Rev. A {\bf 
 31}, 3068 (1985); B. L. Schumaker, Phys. Rev. A {\bf 31}, 3093 (1985).
  %
\bibitem{mdrentcrit} M. D. Reid, quant-ph/0112038.
%
\bibitem{tele}
 A. Furasawa, J. Sorensen, S. Braunstein, C. Fuchs, H. 
Kimble and E. Polzik, Science {\bf 282}, 706 (1998); 
L. Vaidmann, Phys. Rev. A {\bf 49}, 1473 (1994);
 S. Braunstein and H. J. Kimble, Phys. Rev. Lett. {\bf 
80}, 869 (1998); A. Kuzmich and E. S. Polzik, Phys. Rev. Lett. {\bf 
85}, 5639 (2000); I. V. Sokolov, M. I. Kolobov, A. Gatti, L. A. 
Lugiato, quant-ph/0007026.
 %
\bibitem{cryralph} T. C. Ralph, Phys. Rev. A {\bf 61}, 303 (1999); Phys. 
Rev. A {\bf 62} 062306 (2000).
%
\bibitem{cryhill} 
M. Hillery, Phys. Rev. A {\bf 61}, 2309 (1999).
%
\bibitem{cryreid} M. D. Reid, Phys. Rev. 
A{\bf 62}, 062308 (2000).
%
\bibitem{crycerf}
 N. J. Cerf, M. Levy and G. Van Assche, 
Phys. Rev. A {\bf 63} 052311 (2001).
%
\bibitem{cryhjk}S. F. Pereira, Z. Y. Ou and H. J. Kimble, quant-ph/0003094.
%
\bibitem{crynavz} P. Navz, A. Gatti and G. Lugiato, 
quant-ph/0101113; P. Navz, A. Gatti and G. Lugiato, 
to be published. 
%
\bibitem{crysil} Ch. Silberhorn et al, Europe IQEC (2000). 
   %
\bibitem{cryduan}
 L. M. Duan, J. I. Cirac, P. Zoller, Phys. Rev. 
Lett. {\bf 85}, 5643 (2000). 
   %
 \bibitem{numbercry} C. H. Bennett and G. Brassard, in Proceedings of IEEE 
International Conference on Computers, Systems and Signal Processing, 
Bangalore, India (IEEE, New York, 1984), p. 175. 
A. K. Ekert, Phys. Rev. Lett, {\bf 67}, 661 (1991). 
C. H. Bennett, G. Brassard and N. D. Mermin, Phys. Rev. Lett. {\bf 
68}, 557 (1992). A. K. Ekert, J. G. Rarity, P. R. Tapster and G. M. Palma, Phys. Rev. 
Lett {\bf 69}, 1293 (1992). A. K. Ekert, B. Huttner, G. M. Palma and A.  
Peres, Phys. Rev. A {\bf 50}, 1047 (1994). 
%
 \bibitem{Bohm}  D. Bohm, ``Quantum Theory'' (Prentice-Hall, Englewood Cliffs, 
 NJ, 1951). 

\end{references}
\end{document}